\documentclass[aps,prd,reprint,groupedaddress,preprintnumbers]{revtex4-1}
\usepackage{amsmath}
\usepackage{amsfonts} 
\usepackage{amssymb}
\usepackage{graphicx}
\usepackage{hyperref}

\newcommand{\be}{\begin{equation}}
\newcommand{\ee}{\end{equation}}
\newcommand{\bea}{\begin{eqnarray}}
\newcommand{\eea}{\end{eqnarray}}

\begin{document}

\title{Machine Learning Line Bundle Cohomologies of Hypersurfaces in Toric Varieties}
\preprint{MPP-2018-222}

\author{Daniel Klaewer, Lorenz Schlechter}
\affiliation{ Max-Planck-Institut f\"ur Physik (Werner-Heisenberg-Institut),
             F\"ohringer Ring 6,
             80805, M\"unchen, Germany}

\begin{abstract}
Different techniques from machine learning are applied to the problem of computing line bundle cohomologies of (hypersurfaces in) toric varieties. While a naive approach of training a neural network to reproduce the cohomologies fails in the general case, by inspecting the underlying functional form of the data we propose a second approach. The cohomologies depend in a piecewise polynomial way on the line bundle charges. We use unsupervised learning to separate the different polynomial phases. The result is an analytic formula for the cohomologies. This can be turned into an algorithm for computing analytic expressions for arbitrary (hypersurfaces in) toric varieties.
\end{abstract}

\maketitle

\section{Introduction}

The idea of applying concepts from data science to problems naturally appearing in string phenomenology is of course not new. The emergence of the string landscape, the set of effective field theories arising from some consistent string construction, has quickly lead people to consider statistical tools to tackle its enormous size \cite{Douglas:2003um}.

Following early work on genetic algorithms \cite{Abel:2014xta,Allanach:2004my}, with techniques from data science and machine learning recently becoming  important for the solution of many real world problems, there has been an increased interest in applying machine learning wisdom to the exploration of the landscape \cite{He:2017aed,He:2017set,Krefl:2017yox,Ruehle:2017mzq,Carifio:2017bov}\cite{Carifio:2017nyb,Wang:2018rkk,Bull:2018uow,Erbin:2018csv}.

We want to stress here that while e.g. the number of flux vacua is numerically huge (the famous estimated lower bound being $10^{500}$), we are still dealing with a possibly finite and likely countable set whose members can be described by a vector with integral entries. Often the answer to many interesting questions about the vacua can also be described by a set of integers, as is the case for yes/no questions of the type ``Is my vacuum supersymmetric?'' or ``Does my vacuum contain a tachyon'', but also questions such as ``How many generations of SM-fermions does my vacuum contain?''. We want to address the question whether such (complicated) mappings between vectors of integers can be naturally modelled by neural networks (NNs). A particular such questions is:

\emph{``Given a (hypersurface in a) toric variety $X$, what are the ranks $h^\bullet$ of the line bundle cohomology groups $H^\bullet \mathcal({O}_X(D))$, for some toric divisor D?''}

In many cases the answer to this question is provided by the cohomCalg program \cite{cohomCalg:Implementation}, which supports us with data sets on which neural networks can be trained.

As a first approach, we try to directly train a neural network to reproduce the cohomologies. We study first whether this approach can work for the toric ambient spaces and also hypersurfaces therein. The possibility of interpolating and extrapolating the data from a training set is then investigated. This approach is very similar to the one adopted in \cite{Ruehle:2017mzq}, where genetic algorithms were employed to optimise a neural network for regression of line bundle cohomologies.

Our second approach consists of a two step procedure. First we cluster the cohomology data using unsupervised learning. The resulting clusters turn out to have a simple polynomial formula for their cohomologies. The two steps lead to an analytic expression for the rank of the line bundle cohomology groups.

On the way we solve a shortcoming of the cohomCalg algorithm by implementing some of the mappings in the Koszul complex.

After completion of this work we became aware of \cite{Constantin:2018hvl}, which deals with the similar problem of computing line bundle cohomologies in the case of CICYs in products of projective spaces.

\section{Line Bundles on Hypersurfaces in Toric Varieties}

A vast majority of the Calabi-Yau manifolds that are used in string constructions are obtained as complete intersections in toric varieties, the anticanonical hypersurfaces forming a subset of these. Although our techniques are expected to generalise to the case of complete intersections, we will treat only the case of hypersurfaces as a proof of principle.

Toric varieties can be described in many different ways, one of which is the gauged linear sigma model (GLSM) \cite{Witten:1993yc}. The GLSM is an $\mathcal{N}=(2,2)$ SUSY gauge theory in two dimensions, with chiral superfields $x_i$, $i=1,\dots,I$, representing homogeneous coordinates of the toric space. The GLSM features $R$ abelian gauge symmetries, and the charge vectors $Q_i^{(r)}$, $r=1,\dots,R$ encode the weights under $(\mathbb{C}^*)^R$ rescalings of the homogeneous coordinates. Analogous to the case of projective spaces, the resulting toric variety $X$ is then formed as a quotient of $\mathbb{C}^I$ by the homogeneous rescalings, after cutting out a suitable fixed point set $F$
\begin{equation}
	X=\frac{\mathbb{C}^I-F}{(\mathbb{C}^*)^R}\;.
\end{equation}
This fixed point set depends on the choice of the FI parameters in the gauge theory. Solvability of the D-terms will result in the constraint that certain subsets $\mathcal{S}_\alpha$ of the full set of coordinates should not vanish simultaneously
\begin{equation}
	\mathcal{S}_\alpha=\left\{x_{\alpha_1},\dots,x_{\alpha_{|\mathcal{S}_\alpha|}}\right\}\,\qquad,\alpha=1,\dots,N\;.
\end{equation}
The extracted set then takes the form
\begin{equation}
	F=\bigcup\limits_{\alpha=1}^N \left\{x_{\alpha_1}=\dots=x_{\alpha_{|\mathcal{S}_\alpha|}}=0\right\}\;.
\end{equation}
The ring-theoretic way of handling the information in the vanishing set is given by the Stanley-Reisner ideal
\begin{equation}
	\text{SR}=\left<\tilde{\mathcal{S}}_1,\dots,\tilde{\mathcal{S}}_N\right>\;.
\end{equation}
Here the generators $\tilde{\mathcal{S}}_\alpha=\prod_{i=1}^{|\mathcal{S}_\alpha|} x_{\alpha_i}$ are monomials constructed out of the coordinates in the sets $\mathcal{S}_\alpha$.

The homogeneous coordinates of a toric variety provide us with a natural open covering in terms of the sets $U_i=\{x|x_i\neq 0\}$ as well as a set of divisors $D_i=\{x|x_i=0\}$. Due to the equivalence between line bundles and divisors, line bundles on a toric variety take the form of tensor products of the $L_i=\mathcal{O}_X(D_i)$ and their inverses. We can also classify line bundles in terms of their GLSM charges as
\begin{equation}
	L_i=\mathcal{O}_X\left(Q_i^{(1)},\dots,Q_i^{(R)}\right)\;.
\end{equation}
In a toric variety, the anticanonical hypersurface $H=\sum_iD_i$ has vanishing first Chern class and is thus Calabi-Yau. Line bundles $\mathcal{O}_X(D)$ on the ambient space descend to line bundles on this hypersurface $\mathcal{O}_H(D)$. The two are related by an exact sequence of sheaves, the Koszul sequence
\begin{equation}
\label{eq:Koszulseq}
	0\to\mathcal{O}_X(D-H)\stackrel{m}{\to}\mathcal{O}_X(D)\stackrel{res}{\to}\mathcal{O}_H(D)\to 0\;.
\end{equation}
Here $m$ is multiplication with the defining section of $O_X(H)$ of the hypersurface and $res$ is the restriction map to it. Our main interest are the sheaf cohomology groups $H^\bullet(\mathcal{F})$ for the sheaves $\mathcal{F}=\mathcal{O}_X(D),\,\mathcal{O}_H(D)$. In principle the ambient space cohomology can be computed in a brute force way as the \v{C}ech cohomology $\check{H}^\bullet(\mathcal{F},\mathcal{U})$ with respect to the open cover $\mathcal{U}$ defined by the $U_i$.

\section{The cohomCalg Algorithm}
\label{sec:cohomcalg}

A more elegant and fast way to compute the sheaf cohomology is given by the cohomCalg algorithm, which has been conjectured in \cite{CohomOfLineBundles:Algorithm}, proven in \cite{Rahn:2010fm} and implemented in \cite{cohomCalg:Implementation}. The algorithm gives generators of the cohomology groups in terms of \emph{rationoms}, which are just monomials of the form
\begin{equation}
\label{eq:rationom}
	\frac{T(\vec{x})}{(\prod y_i)\cdot W(\vec{y})}\;,
\end{equation}
where the vectors $\vec{x},\vec{y}$ refer to a splitting of the homogeneous coordinates as follows. The power set of the Stanley-Reisner ideal\footnote{Here power set means the set of all possible unions of generators of the ideal.} is decomposed into its k-element subsets as
\begin{equation}
	P(\text{SR})=\bigcup\limits_{k=0}^{|\text{SR}|}P_k(\text{SR})\;.
\end{equation}
One defines index-sets $A=\{\alpha_1,\dots,\alpha_k\}\subset\{1,\dots,|\text{SR}|\}$ which allow us to label the elements of the sets $P_k(\text{SR})$ as $\mathcal{P}^k_A=\{\tilde{\mathcal{S}}_{\alpha_1},\dots,\tilde{\mathcal{S}}_{\alpha_k}\}$. For a given $\mathcal{P}^k_A$, the union of all its associated $\mathcal{S}_{\alpha_i}$ is denoted as
\begin{equation}
	\mathcal{Q}_A^k=\bigcup\limits_{i=1}^k\mathcal{S}_{\alpha_i}\;,
\end{equation}
which is just the collection of all coordinates that appear in the set $\mathcal{P}^k_A$. To this set, a degree $N^k_A$ is assigned:
\begin{equation}
	N^k_A=\left|\mathcal{Q}^k_A\right|-k\,.
\end{equation}

For a given $\mathcal{Q}=\mathcal{Q}^k_A$ the variables $\vec{y}$ that appear in the denominator of the rationom \eqref{eq:rationom} are now defined to be those that are contained in $\mathcal{Q}$, whereas the $\vec{x}$ coordinates are taken from the complement. For this given $\mathcal{Q}$ we can now construct all possible rationoms that match the GLSM charge of the divisor $D$ that defines the line bundle $\mathcal{O}_X(D)$. Each rationom contributes a generator of the cohomology group $H^N(X,\mathcal{O}_X(D))$, with $N=N^k_A$.

In some cases a single rationom will contribute multiple generators to the cohomology. This is associated with the calculation of a certain remnant cohomology, which has been clarified in \cite{Rahn:2010fm}. Although these multiplicities are implemented in the cohomCalg program, this complication will not appear in the examples that we study.

Once the sheaf cohomology of $X$ is computed, one can use the fact that the short exact sequence of sheaves \eqref{eq:Koszulseq} induces a long exact sequence of cohomology groups
\begin{equation}
\begin{aligned}
	\cdots&\stackrel{\delta}{\to} H^i(\mathcal{O}_X(D-H))\stackrel{m_*}{\to}H^i(\mathcal{O}_X(D))\stackrel{res_*}{\to}\\
	&\stackrel{res_*}{\to}H^i(\mathcal{O}_H(D))\stackrel{\delta}{\to}H^{i+1}(\mathcal{O}_X(D-H))\stackrel{m_*}{\to}\cdots
\end{aligned}\;,
\end{equation}
where $\delta$ is the connecting homomorphism, in order to deduce the sheaf cohomology $H^\bullet(H,\mathcal{O}_H(D))$ on the hypersurface.

The reference implementation of the cohomCalg algorithm \cite{cohomCalg:Implementation} does not implement the maps in the Koszul-sequence and hence relies on the exactness of the sequence in order to derive the ranks of the cohomology groups. This works by first cutting the long sequence into shorter sequences at locations where zeros occur and then using the fact that for an exact sequence
\begin{equation}
	0\to G_1\to\dots\to G_n\to0
\end{equation}
the ranks satisfy $\sum_{j=1}^n (-1)^j \text{rk}(G_j)=0$.

The above approach works as long as there are sufficiently many zeros in the sequence. In order to train our classifiers we need the cohomology ranks of \emph{all} line bundles corresponding to a certain interval $[-\delta,+\delta]$ in charge space. Generically only some of those ranks can be solved by the cohomCalg program, whereas a large portion is left undetermined.

We improve the algorithm by cutting the sequences also at the multiplication maps $m_*$ as
\begin{equation}
\begin{aligned}
	\cdots\to H^i(\mathcal{O}_X(D-H))&\stackrel{m_*}{\to}\text{image}(m_*)\to 0\\
	0\to\text{coker}(m_*)&\stackrel{res_*}{\to}H^i(\mathcal{O}_H(D))\to\cdots
\end{aligned}\;.
\end{equation}

The price for inserting an additional zero is now that we have to compute the (rank of the) image of the map $m_*$. The induced map $m_*$ on the cohomologies is realised in this setting by multiplication of the rationom representatives of the cohomology generators with the defining section $s\in\Gamma(X,\mathcal{O}_X(H))$ of the hypersurface. If a resulting monomial is not contained in the set of rationoms spanning the codomain, it is equivalent to zero in cohomology.

For definitiveness we will always consider the hypersurface to be at the large complex structure point of its moduli space. This means that the map $m$ is just multiplication by the monomial $x_1\cdots x_I$. 

In all cases studied the resulting exact sequences could now be solved for the cohomologies on the hypersurface. If this would have not been the case, we could have also introduced additional cuts at the restriction maps.

The procedure suggests a natural generalization to the case of CICYs in toric varieties for which there exists a similar Koszul sequence, the mappings of which can be implemented in an analogous way. We leave an implementation of this more general case for future work.

Let us outline the calculation in an example. The anticanonical hypersurface in $\mathbb{P}^3_{1112}$ is a K3 surface. The toric resolution of this is described by the charge vector
\begin{equation}
	Q=\bordermatrix{
		&x_1&x_2&x_3&x_4&x_5\cr
		&1&1&1&0&2\cr
		&0&0&0&1&1
	}\;,
\end{equation}
with Stanley-Reisner ideal $\text{SR}=\left<x_1 x_2 x_3, x_4 x_5\right>$. We want to compute the image of the map
\begin{equation}
	H^1\left(\mathcal{O}(-3,-4)\right)\stackrel{m_*}{\to}H^1\left(\mathcal{O}(2,-2)\right)\;,
\end{equation}
where we have introduced a basis $D_1=\{x_1=0\}\sim\{x_2=0\}\sim\{x_3=0\}$ and $D_2=\{x_4=0\}$ of divisors such that $\{x_5=0\}=2D_1+D_2$ and use the corresponding dual basis for the first cohomology.
Using the cohomCalg algorithm we determine the generators of both cohomology groups to be
\begin{equation}
	\begin{aligned}
		H^1\left(\mathcal{O}(-3,-4)\right)&=\left<\frac{(\text{deg } 1 \text{ in }x_{1,2,3})}{x_4^2x_5^2},\frac{(\text{deg } 3 \text{ in }x_{1,2,3})}{x_4x_5^3}\right>\\
		H^1\left(\mathcal{O}(2,-2)\right)&=\left<\frac{(\text{deg } 4 \text{ in }x_{1,2,3})}{x_4x_5}\right>\;.
	\end{aligned}\;
\end{equation}
Under the map $m=\cdot\prod_i x_i$ it is clear that only the first class of generators with denominator $x_4^2 x_5^2$ will be mapped to rationoms that exist in $H^1\left(\mathcal{O}(2,-2)\right)$. The second class of generators with denominator $x_4x_5^3$ is mapped to monomials without $x_4$ in the denominator, which do not have the correct singularity structure to be members of $H^1\left(\mathcal{O}(2,-2)\right)$ and hence are cohomologous to zero. As a result we find that $\text{rk}(\text{im}(m_*))=3$.

For an arbitrary point in the complex structure moduli space the map $m_*$ will of course be more complicated. The polynomials that result from multiplication of the rationoms in $H^1\left(\mathcal{O}(-3,-4)\right)$ with the defining polynomial of the hypersurface will have to be reduced modulo the rationoms in the target cohomology. While this is straightforward to implement, it is computationally more expensive and we restrict to the large complex structure point to illustrate our methods.

\section{Machine Learning Cohomologies}
The aim of this paper is to examine the possible application of neural networks in the computation of line bundle cohomologies of toric varieties and hypersurfaces therein. There are different possible approaches. In \cite{Ruehle:2017mzq} genetic algorithms were used to evolve neural networks which were then used to perform a regression on the map between the line bundle charges and cohomologies. The resulting NNs reproduced the cohomology ranks with 72\%/83\% accuracy after training. On the other hand the authors of \cite{ Bull:2018uow} used a classification neural network to learn the Hodge numbers of the Kreuzer-Skarke list and achieved a $80\%$ validation rate in predicting the cohomologies.  They also used a regressional neural network to solve the same problem with worse results. While these approaches work in their respective areas of application, they require large data sets and fail at the extrapolation of large numbers.

\subsection{Neural Networks for Classification}
A neural network for a classification problem maps an input vector via several hidden layers, which normally are taken to be ReLU, to a fixed number of output nodes representing the classes. The output is normalised to sum up to $1$ and interpreted as a probability and this is typically implemented by applying a softmax layer. The prediction is the class of highest probability. The loss function has to be proportional to the deviation from the true result and for classification networks often is taken to be the cross entropy. This approach has the severe limitation that one has to a priori fix the possible outcomes, as every possible value of the $h^i$ has a corresponding node. The authors of \cite{Bull:2018uow} avoided this problem by declaring all  $h^i>50$ as large and do not try to classify these. In the examples we will be discussing the ranks can become arbitrarily large and this classification no longer makes sense. While this approach is easy to use, the rather bad results and limitations to very small ranks render it uninteresting.

\subsection{Neural Networks for Regression}
Another approach is a regressional neural network. Here the input vector is again mapped by several hidden ReLU layers to an output vector. This time the output vector is not normalised but takes any value in $\mathbb{R}^n$ and is interpreted as the ranks by rounding to the nearest integer. The loss function for training is taken to be the mean squared error of the prediction compared to the real ranks. This approach does not put a hard upper bound on possible ranks, but the precision of the result is limited by the number of neurons and the floating point precision used. Most standard implementations of NNs use only single precision, resulting in a precision of the ranks of $10^{-6}$. Thus if the ranks exceed $10^6$, the error becomes order one and the NN predicts wrong numbers. 

Moreover, the NN only learns an interpolation of the given data. Therefore, if one trains the network on a data set where the entries of the charge vector are in a certain range, the predictions outside of this range are unreliable.

To illustrate these findings, we take the ambient space $dP_3$ and the hypersurface $\mathbb{P}_{11222}[8]$. We randomly generated $50000$ data points with line bundle charges in the range $[-50,50]$. In the case of $dP_3$, the cohomologies can be learned by a NN consisting of 3 hidden ReLU layers with 500 neurons each to a precision of $99.85\%$ within one hour. In the case of $\mathbb{P}_{11222}[8]$, this approach fails. Even large nets produce only $0.1\%$ correct results. The reasons are that the ranks in this example already exceed $2\cdot10^7$ and the high non-linearity of the problem. Sophisticated preprocessing of the data increased this to $55\%$ accuracy after 10 minutes of training, which is still not satisfactory. Thus for these kind of problems another approach is needed.

\section{An Algorithm to Determine Analytic Formulas}

The algorithm described in section \ref{sec:cohomcalg} allows the determination of the ranks of the cohomology groups for given values of the line bundle charges. In this section an algorithm using unsupervised learning is presented which allows the identification of analytic expressions. 

First a data set $S$ of the cohomologies is calculated for all values of the line bundle charges $\vec{m}$ satisfying $|m_i|\le a \;\forall i$ for a fixed value of $a$. Tests have shown that $a=25$ is sufficient for the algorithm to find the analytic formulas.

The algorithm uses the observation that the $h^i$ have a distinct phase structure. In the interior of one phase the $h^i$ are polynomial functions of the line bundles of maximal degree $d$, where $d$ is the dimension of the variety. If one can identify the phase structure, it is then easy to perform a polynomial fit. This represents a classification problem. As one a priori does not know the phase structure, unsupervised learning has to be applied.

In unsupervised learning one faces the task to group data points into different sets without specifying any conditions. This leads to a clustering of similar data. The only input is the data to classify and the maximal number of sets to be used. We applied the pre-implemented ClusterClassify function of Mathematica 11.3 with 200 classes and ``Quality" as optimization goal as well as ``KMeans" as the method to generate the classifiers and the LinearModelFit function for the polynomial fits.

In the interior of one phase, the $d$-th derivatives of the $h^i$ with respect to $\vec{m}$ are constant and the $(d+1)$-th derivatives vanish. As the $h^i$ are only defined for integer $\vec{m}$, the data forms a lattice. The derivatives are therefore calculated using the central difference scheme with a lattice spacing of one. This leads to a non-vanishing $(d+1)$-th derivative exactly at the phase boundaries. The first step is to remove the boundaries out of the data set $S$. To do so a cluster classifier with a very large number of classes is trained on the data set
\begin{equation}
\left\{\vec{m}\;,\;{\partial^{d+1} h^i\over \partial^{d+1} m_1}\;,\;{\partial^{d+1} h^i\over \partial^{d} m_1\partial m_2}\;,.....\;,\;{\partial^{d+1} h^i\over \partial^{d+1} m_R}\right\}\;,
\end{equation}
where $i=0,\dots,d$ runs over all cohomology groups. This set takes for a point inside a phase the form
\begin{equation}
\{\vec{m},0,0,0,.....,0\}\;
\end{equation}
and for a point at a phase boundary at least one of the latter entries is non-vanishing. This leads to a classification where all data points which lie in the interior of a phase are classified into one set and various sets of boundary points. For large enough line bundle charges the interior will always be the largest set. The boundaries are simply thrown away. Tests show that the classification works better for a small dimensional space. The number of partial derivatives increases with the degree $d$ and the number of line bundle charges. Therefore this step was divided into several classification steps. First one trains one classifier on a subset of the derivatives of degree $d+1$ and removes the boundary.  Then a second classifier is trained on the next subset and so on. As the training of one classifier takes only seconds, this is not a huge performance loss but drastically improves the result. In the examples presented in this paper we used a splitting into two randomly chosen subsets of equal size.

With the remaining points forming the interior of the phases  the set
\begin{equation}
S_3=\left\{\vec{m}\;,\;{\partial^{d} h^i\over \partial^{d} m_1}\;,\;{\partial^{d} h^i\over \partial^{d} m_1\partial m_2}\;,.....\;,\;{\partial^{d} h^i\over \partial^{d} m_R}\right\}
\end{equation}
is formed and a second classifier trained on this set. The set $S_3$ is, in contrary to the original data set S, not connected in the $\vec{m}$, which improves the classification and is the reason for the two step procedure. This now classifies the phase structure of the problem. The number of allowed classes is again taken to be very large. While it can happen that one phase is grouped into two classes, this does not pose any problem as in this case the polynomials obtained will agree and the phases can be merged later on.

The final step is to perform the polynomial fit on each set and each $h^i$. Sets with identical polynomials for all $h^i$ are then merged. This concludes the algorithm. To summarise:

\begin{enumerate}
\item Calculate a set of data points using the extended cohomCalg.
\item Determine the $(d+1)$-th derivatives of these points.
\item Classify the data using these derivatives.
\item Determine the $d$-th derivatives of the remaining data points.
\item Classify the data using these derivatives.
\item Perform a polynomial fit of degree $d$ on each set for each $h^i$.
\item Merge sets with identical polynomials.
\end{enumerate}

We note that this algorithm requires no input besides the geometric data describing the variety and can therefore be completely automatised. The only thing which has to be done by hand is to extract the boundaries of the phases, as the classifier encodes them not in closed form. This is quite tedious, but for practical purposes one does not need the functions. One can use the classifier to identify in which phase a given $\vec{m}$ lies and apply the polynomial of this phase. For convenience we added the phase boundaries in the tables. 

As a non-trivial test of the procedure we calculated the Euler characteristic of the examples by summing up the polynomials and compare them to the Euler characteristic as obtained from the Hirzebruch-Riemann-Roch theorem. The two expressions agree in all examples and phases.

In the following sections this algorithm is applied to some examples.

\section{Line Bundles on Toric Varieties}

We start with an example where the analytic expressions are well known, the del Pezzo surface $dP_1$. This provides on one hand an easy method to cross-check the results and on the other hand is an easy example with only 3 phases. 

\begin{figure}[h!]
	\centering
	\includegraphics[scale=0.27]{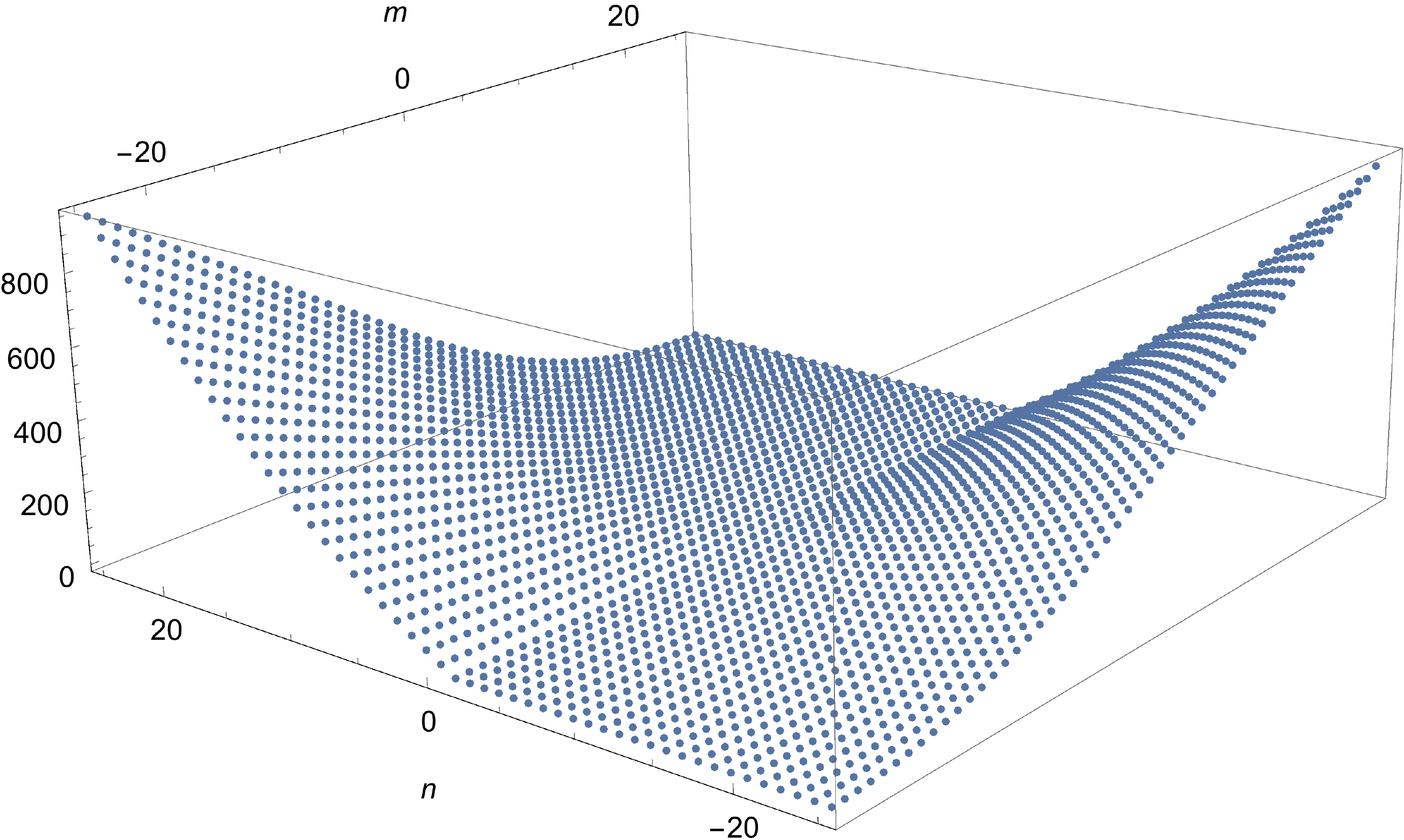}
	\caption{$h^1\left(\mathcal{O}(m,n)\right)$ of $dP_1$.}
	\label{dp1fig1} 
	\end{figure}

\begin{figure}[h!]
	\centering
	\includegraphics[scale=0.27]{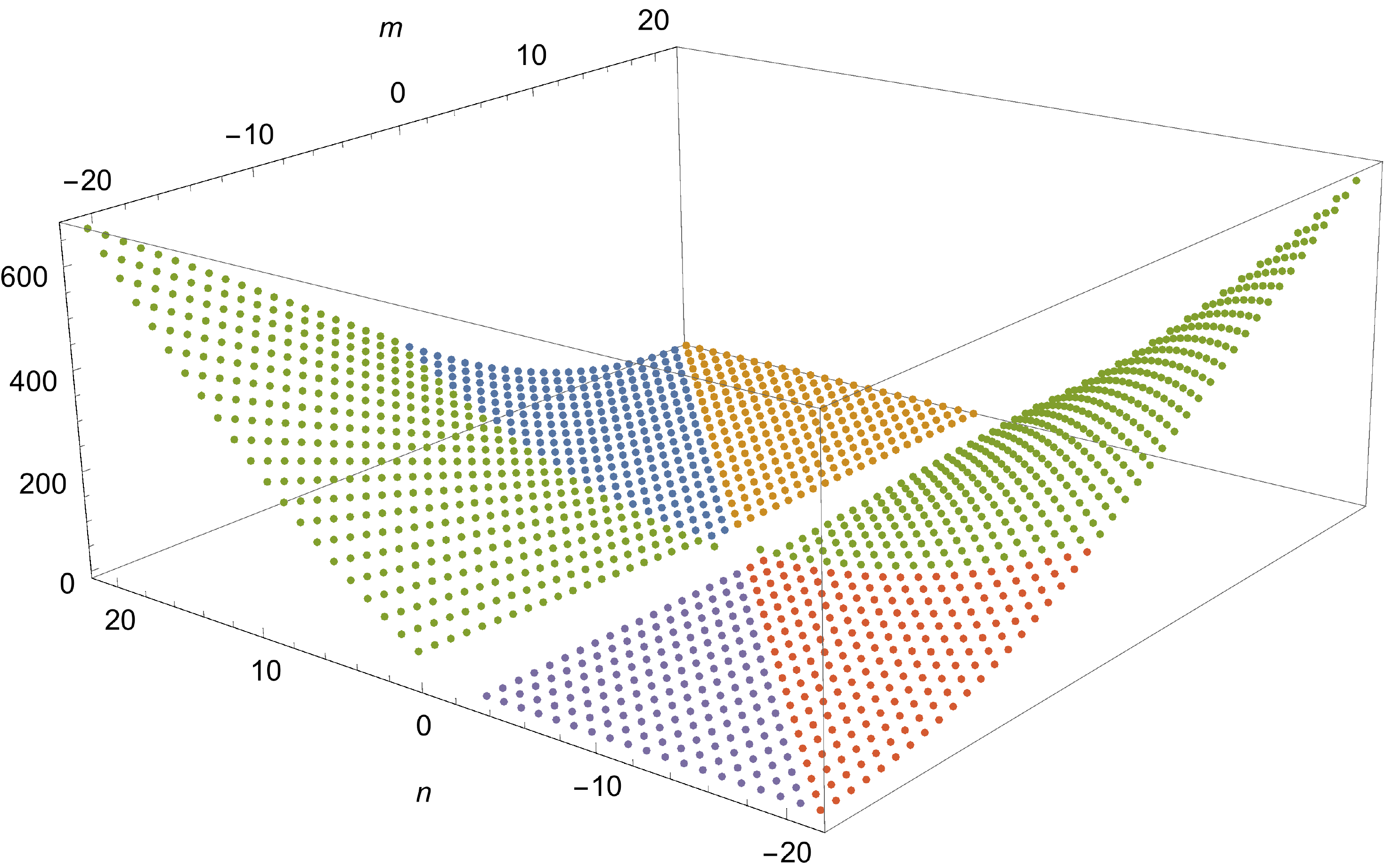}
	\caption{Classification result for $h^1\left(\mathcal{O}(m,n)\right)$ of $dP_1$.}
	\label{dp1fig2} 
	\end{figure}

Using cohomCalg, we generate a data set of the cohomology ranks with the line bundle charges in the range $a=[-25,25]$. These are shown in figure \ref{dp1fig1}. The application of the unsupervised learning on the third derivatives cuts out two phase-boundaries where the underlying function describing the ranks is non-differentiable. The second cluster analysis then classifies the remaining points using the second derivatives into $6$ phases, three pairs of which have identical polynomials for $h^1$. The result is shown in figure \ref{dp1fig2}. 

Fitting a polynomial of degree $2$  to the ranks in each of these phases results in the polynomials listed in table \ref{tabledP1}. These agree with the known analytic expressions, see e.g. \cite{CohomOfLineBundles:Algorithm}. 

\begin{table}
\begin{tabular}{|c|c|}
\hline
Phase&Polynomial\\
\hline
$\begin{aligned}(n\leq -2&\land m\geq 0)\\\lor (n\geq 0&\land m\leq -3)\end{aligned}$&$-1-m-{n\over 2}-mn+{n^2\over 2}$\\
\hline
$\begin{aligned}(n\leq -2&\land n+1\leq m<0)\\\lor (n\geq 0&\land -3<m\leq n-2)\end{aligned}$&${m\over 2}+{m^2\over 2}-{n\over 2}-mn+{n^2\over2}$\\
\hline
else&$0$\\
\hline
\end{tabular}
\caption{Polynomials for $h^1\left(\mathcal{O}(m,n)\right)$ in the case of $dP_1$.}
\label{tabledP1}
\end{table}

\section{Line Bundles on Hypersurfaces}
\begin{figure}[h!] 
	\centering
	\includegraphics[scale=0.27]{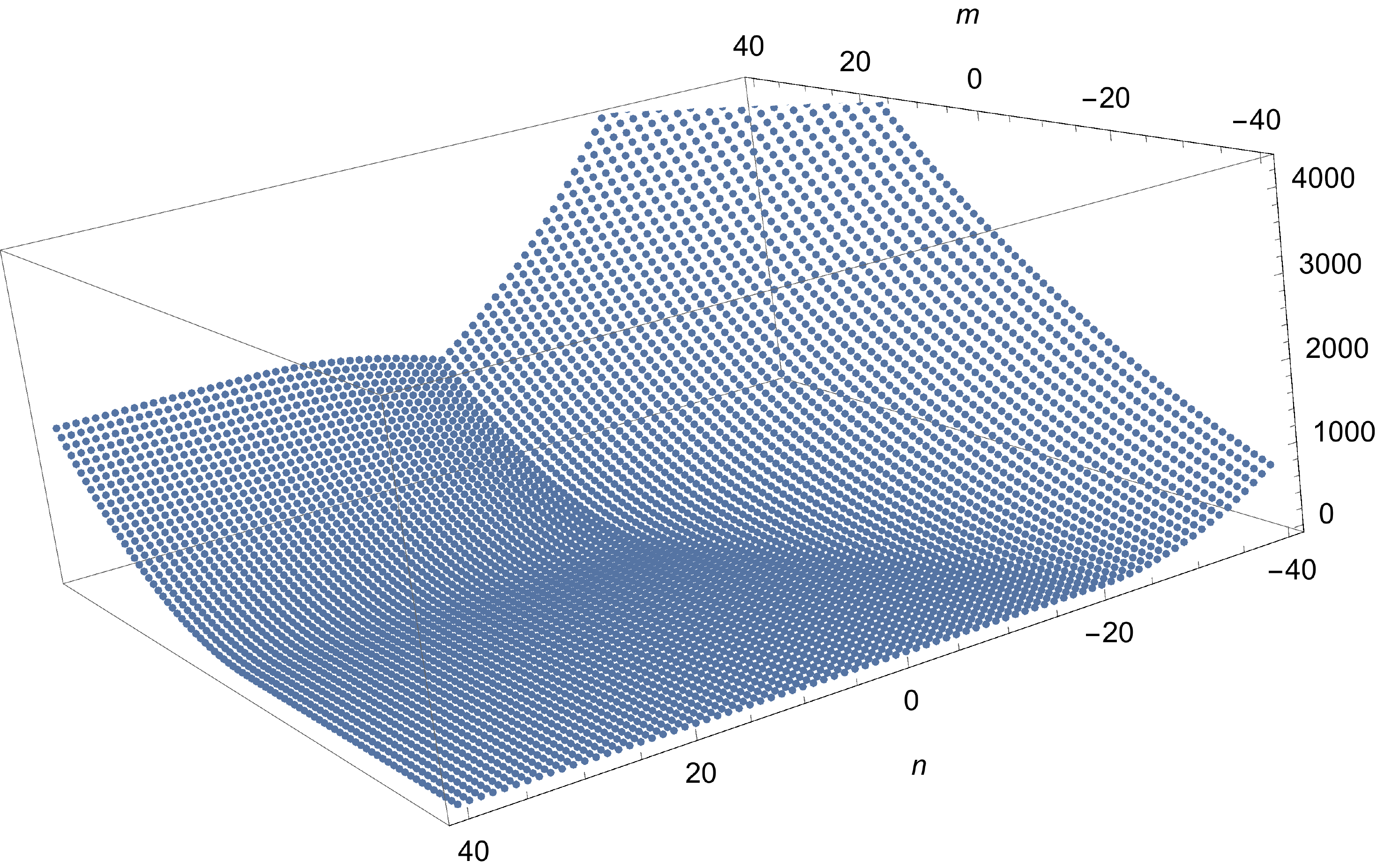}
	\caption{$h^0\left(\mathcal{O}(m,n)\right)$ of $\mathbb{P}^3_{1112}[5]$.}
	\label{K3fig1}
	\end{figure}

We now turn to the more complicated problem of finding analytic expressions for line bundle cohomologies of hypersurfaces in toric varieties.  As an example for a hypersurface we take the K3 space $\mathbb{P}^3_{1112}[5]$. This hypersurface has two line bundle charges, so that $\vec{m}=(m,n)$. The expected degree of the polynomials is $d=2$. Figure \ref{K3fig1} shows the ranks of the zeroth cohomology for different values of $m$ and $n$.

\begin{figure}[h!]
	\centering
	\includegraphics[scale=0.27]{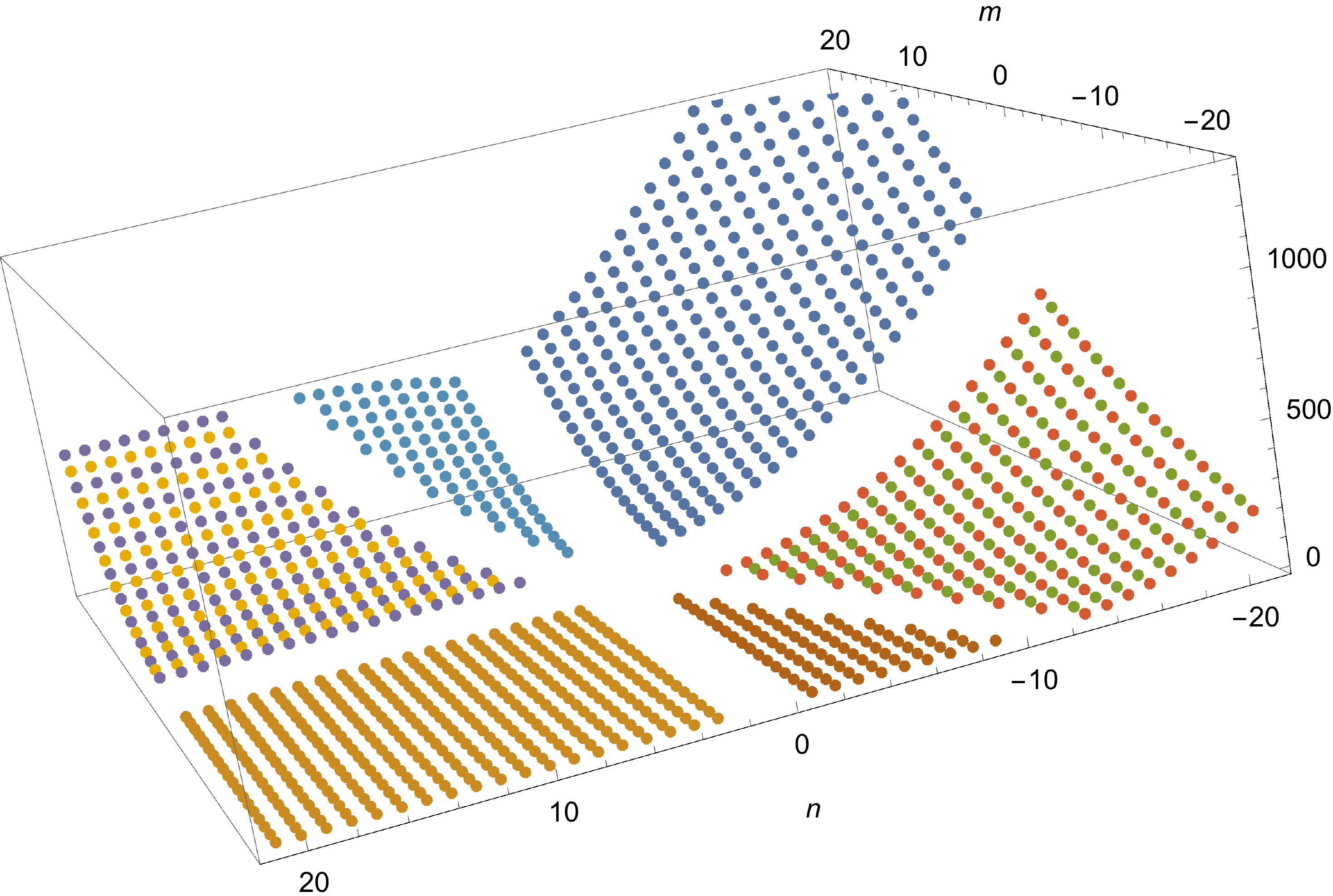}
	\caption{$h^0\left(\mathcal{O}(m,n)\right)$ of $\mathbb{P}^3_{1112}[5]$ separated into phases.}
	\label{K3fig2} 
	\end{figure}

At first glance this seems to consist of 3 phases. But applying the algorithm described in the last section reveals that there are actually 6 phases. Figure \ref{K3fig2} shows the result of the second classification. The fitted polynomials can be found in table \ref{tablep1112}. One nicely sees the cut boundaries and phases. Also the separation between the orange and brown phase seems redundant from the point of view of $h^0$, but is necessary because of the higher cohomology groups. Especially interesting is the subdivision in the yellow/purple and red/green phases into even and odd $n$, which are also described by different polynomials. The phase structure thus is not only defined by some linear functions of $m$ and $n$. If one tried a polynomial fit in the whole of these phases instead of separating into even/odd one would not obtain rational coefficients. E.g. in the yellow/purple phase the polynomials are $\frac{5 m^2}{4}+2$ for $n$ even and $\frac{5 m^2}{4}+\frac{7}{4}$ for $n$ odd. If one mixes these phases, the interpolating polynomial obtained is  $1.80407+0.0131771\, n+1.24945\, n^2$, which does obviously not reproduce any of the cohomologies correctly and cannot be extrapolated.

\begin{table}
\begin{tabular}{|c|c|}
\hline
Phase&Polynomial\\
\hline
$m<0,n>{m\over 2}$&$0$\\
\hline
$m<0,n<{m\over 2}$&$\frac{m^2}{2}-2 m n-\frac{3 m}{2}+2 n^2+3 n+1$\\
\hline
$m>0,n>{m\over 2}, m \text{ even}$&$\frac{5 m^2}{4}+2$\\
\hline
$m>0,n>{m\over 2}, m \text{ odd}$&$\frac{5 m^2}{4}+\frac{7}{4}$\\
\hline
$m>0,0<n<{m\over 2}$&$m^2+m n-n^2+2$\\
\hline
$m>0,n<0$&$m^2-2 m n-3 m+2 n^2+3 n+2$\\
\hline
\end{tabular}
\caption{Polynomials for $h^0\left(\mathcal{O}(m,n)\right)$ in the case of $\mathbb{P}^3_{1112}[5]$.}
\label{tablep1112}
\end{table}

\begin{figure}[h!]
	\centering
	\includegraphics[scale=0.27]{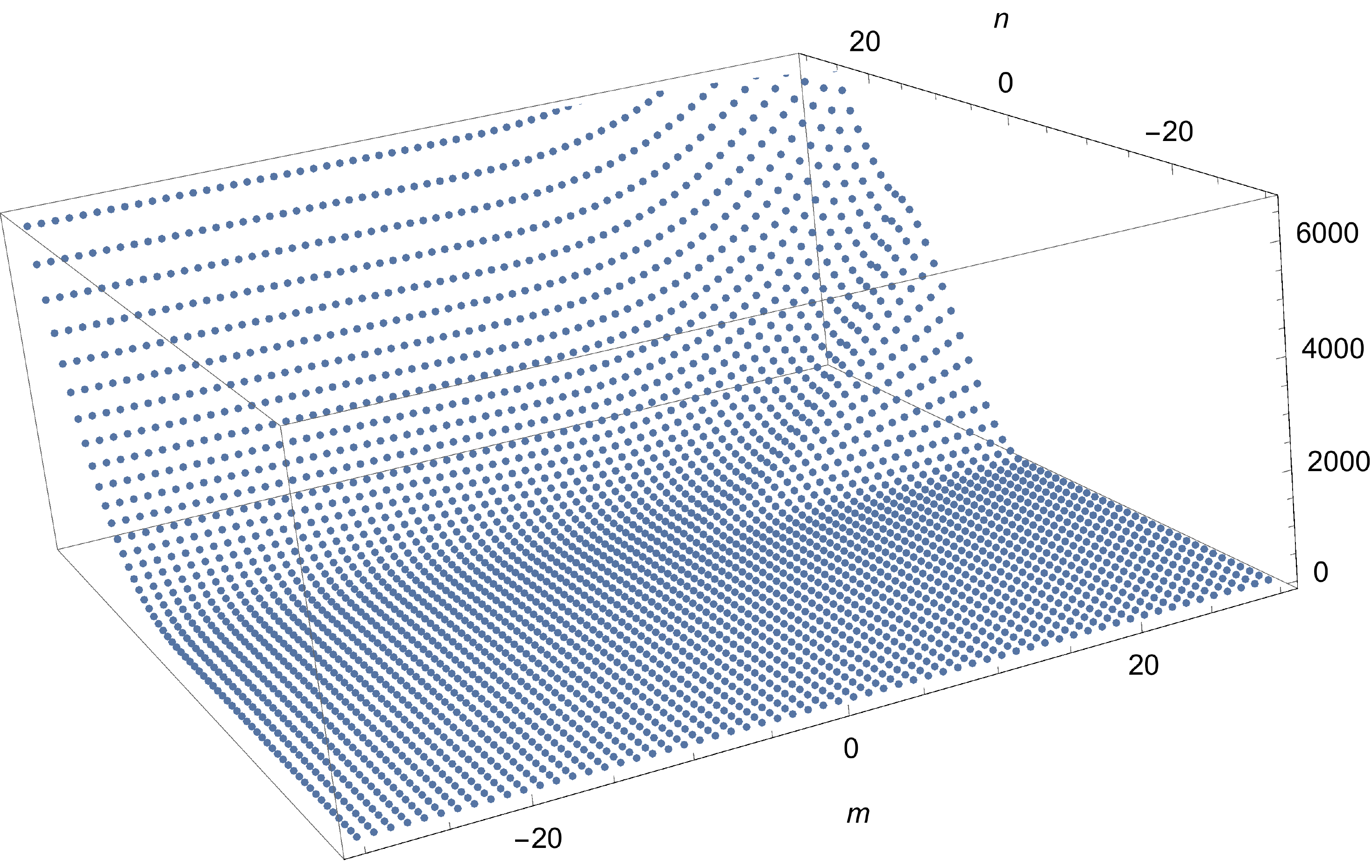}
	\caption{$h^0\left(\mathcal{O}(m,n)\right)$ of $\mathbb{P}^4_{11222}[8]$.}
	\label{P11222fig1} 
	\end{figure}

\begin{figure}[h!]
	\centering
	\includegraphics[scale=0.46]{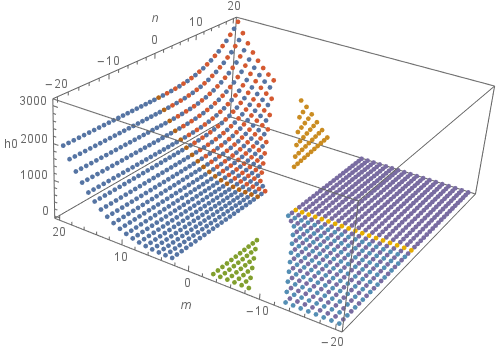}
	\caption{Classification result for $h^0\left(\mathcal{O}(m,n)\right)$ of $\mathbb{P}^4_{11222}[8]$.}
	\label{P11222fig2} 
	\end{figure}

Another interesting example is the octic $\mathbb{P}^4_{11222}[8]$. Here we expect the polynomials to be of degree $d=3$. Figures \ref{P11222fig1} and \ref{P11222fig2} show again the input data for $h^0$ and the result after classification.
The resulting polynomials for $h^0$ are listed in table \ref{tablep11222}.
\begin{table}
\begin{tabular}{|c|c|}
\hline
Phase&Polynomial\\
\hline
$m<0,n\in \mathbb{Z}$&$0$\\
\hline
$m>0,n<0$&$\frac{m^3}{3}-2 m^2+\frac{11
   m}{3}-1$\\
\hline
$m>0,n>{m\over 2}$&$-\frac{8 m^3}{3}+2 m^2 n+\frac{2 m}{3}+2 n$\\
\hline
$m>0,0<n<{m\over 2}$, $m$ even&$\begin{aligned}
	&\tfrac{m^3}{3}-2 m^2+\tfrac{11 m}{3}+\tfrac{n^3}{8}+\tfrac{3 n^2}{8}\\&+\tfrac{5 n}{4}-1
\end{aligned}$\\
\hline
$m>0,0<n<{m\over 2}$, $m$ odd&$\begin{aligned}
	&\tfrac{m^3}{3}-2 m^2+\tfrac{11 m}{3}+\tfrac{n^3}{8}+\tfrac{3 n^2}{8}\\&+\tfrac{7 n}{8}-\tfrac{11}{8}
\end{aligned}$\\
\hline
\end{tabular}
\caption{Polynomials for $h^0\left(\mathcal{O}(m,n)\right)$ in the case of $\mathbb{P}^4_{11222}[8]$.}
\label{tablep11222}
\end{table}
We note that the only disadvantage of this procedure is that the boundaries are cut out and it is not possible to determine the value at the boundaries itself, which is reflected in only $>$ statements in the table instead of $\geq$. But as these are only a limited number of points one can simply compare these with the results from cohomCalg. The tables for the other cohomology groups can be found in appendix \ref{app:A}.

\section{Discussion}

We have presented a method for generating analytic expressions for all line bundle cohomology ranks of toric varieties or hypersurfaces therein. The algorithm takes as an input the toric data in form of GLSM charges and the Stanley-Reisner ideal. For the case of hypersurfaces we also need to specify a point in the complex structure moduli space in the form of a polynomial that defines a section of $\mathcal{O}_X(H)$ and hence a specific hypersurface. For demonstrative purposes we calculated at the large complex structure point but the method carries over to other generic and special points in the moduli space.

The output is a classifier that separates the space of line bundles into different phases, such that within a phase each cohomology is described by a single polynomial in the line bundle charges. Since the polynomials have coefficients in $\mathbb{Q}$ the result can be considered exact and we obtain a formula for all of the line bundles. As a cross-check we see that the alternating sum of polynomials in each phase reproduces the Euler characteristic as calculated from the Hirzebruch-Riemann-Roch theorem.

It was crucial to realise that we understand the local structure of the data and the problem of patching this to obtain the global structure could be broken down to a simple classification problem.

We expect that our methods carry over to similar problems of this type. For example the case of line bundles on complete intersections in toric varieties should be completely analogous. We leave the interesting case of vector bundles of higher rank in the form of monad bundles for future work.

\section*{Acknowledgements}

We are indebted to Ralph Blumenhagen for discussions about line bundle cohomologies and machine learning which initiated this project as well as contributions in the early stages. We are also grateful to Harold Erbin for illuminating conversations about neural networks.

\bibliography{MLCohom}{}

\clearpage

\appendix

\section{Line Bundle Cohomologies}
\label{app:A}
\begin{table}[h!]
\begin{tabular}{|c|p{5cm}|p{5cm}|p{5cm}|}
\hline
Phase&$h^0$&$h^1$&$h^2$\\
\hline
$I$&0&0&$-n^2+n m+m^2+2$\\
\hline
$II$&$-n^2+n m+m^2+2$&0&0\\
\hline
$III$&$\frac{5 m^2}{4}+\frac{7}{4}$&$3 n^2-3 n m-3 n+\frac{3 m^2}{4}+\frac{3 m}{2}+\frac{3}{4}$&$2 n^2-2 n m-3 n+\frac{m^2}{2}+\frac{3 m}{2}+1$\\
\hline
$IV$&$2 n^2-2 n m+3 n+\frac{m^2}{2}-\frac{3 m}{2}+1$&$3 n^2-3 n m+3 n+\frac{3 m^2}{4}-\frac{3 m}{2}+\frac{3}{4}$&$\frac{5 m^2}{4}+\frac{7}{4}$\\
\hline
$V$&$2 n^2-2 n m+3 n+\frac{m^2}{2}-\frac{3 m}{2}+1$&$3 n^2-3 n m+3 n+\frac{3 m^2}{4}-\frac{3 m}{2}+1$&$\frac{5 m^2}{4}+2$\\

\hline
$VI$&$\frac{5 m^2}{4}+2$&$3 n^2-3 n m-3 n+\frac{3 m^2}{4}+\frac{3 m}{2}+1$&$2 n^2-2 n m-3 n+\frac{m^2}{2}+\frac{3 m}{2}+1$\\

\hline
$VII$&$0$&$3 n^2-3 n m-3 n+3 m$&$2 n^2-2 n m-3 n+m^2+3 m+2$\\
\hline
$VIII$&$2 n^2-2 n m+3 n+m^2-3 m+2$&$3 n^2-3 n m+3 n-3 m$&0\\

\hline
\end{tabular}
\caption{Polynomials for all $h^i$ in the case of $\mathbb{P}^3_{1112}[5]$.}
\end{table}
\begin{table}
\begin{tabular}{|c|p{4cm}|p{4cm}|p{4cm}|p{4cm}|}
\hline
Phase&$h^0$&$h^1$&$h^2$&$h^3$\\
\hline
$I$&0&0&0&$\frac{8 m^3}{3}-2 m^2 n-\frac{2 m}{3}-2 n$\\
\hline
$II$&$-\frac{8 m^3}{3}+2 m^2 n+\frac{2 m}{3}+2 n$&0&0&0\\
\hline
$III$&$\frac{m^3}{3}-2 m^2+\frac{11 m}{3}+\frac{n^3}{8}+\frac{3 n^2}{8}+\frac{5 n}{4}-1$&$-1 + 3 m - 2 m^2 + 3 m^3 - (3 n)/4 - 2 m^2 n + (3 n^2)/8 + n^3/8$&0&0\\
\hline
$IV$&0&0&$-3 m^3+2 m^2 n-2 m^2-3 m-\frac{n^3}{8}+\frac{3 n^2}{8}+\frac{3 n}{4}-1$&$-1 - (11 m)/3 - 2 m^2 - m^3/3 - (5 n)/4 + (3 n^2)/8 - n^3/8$\\
\hline
$V$&$\frac{m^3}{3}-2 m^2+\frac{11 m}{3}-2$&$3 m^3-2 m^2 n-2 m^2+3 m-2 n-2$&0&0\\
\hline
$VI$&$\frac{m^3}{3}-2 m^2+\frac{11 m}{3}+\frac{n^3}{8}+\frac{3 n^2}{8}+\frac{7 n}{8}-\frac{11}{8}$&$3 m^3-2 m^2 n-2 m^2+3 m+\frac{n^3}{8}+\frac{3 n^2}{8}-\frac{9 n}{8}-\frac{11}{8}$&0&0\\
\hline
$VII$&0&0&$-3 m^3+2 m^2 n-2 m^2-3 m-\frac{n^3}{8}+\frac{3 n^2}{8}+\frac{9 n}{8}-\frac{11}{8}$&$-\frac{m^3}{3}-2 m^2-\frac{11 m}{3}-\frac{n^3}{8}+\frac{3 n^2}{8}-\frac{7 n}{8}-\frac{11}{8}$\\
\hline
$VIII$&0&0&$-3 m^3+2 m^2 n-2 m^2-3 m+2 n-2$&$-\frac{m^3}{3}-2 m^2-\frac{11 m}{3}-2$\\
\hline
\end{tabular}
\caption{Polynomials for all $h^i$ in the case of $\mathbb{P}^4_{11222}[8]$.}
\end{table}
\end{document}